\documentclass[12pt,preprint]{aastex}

\def\lesssim{\mathrel{\hbox{\rlap{\hbox{\lower4pt\hbox{$\sim$}}}\hbox{$<$}}}}
\def\gtrsim{\mathrel{\hbox{\rlap{\hbox{\lower4pt\hbox{$\sim$}}}\hbox{$>$}}}}

\shorttitle{Magnetically-Supported Protostellar Disks}
\shortauthors{Li}

\begin{document}

\title{Self-Gravitating Magnetically-Supported Protostellar Disks and 
	the Formation of Substellar Companions} 

\author{Zhi-Yun Li}
\affil{Astronomy Department, University of Virginia, P.O.Box 3818,
Charlottesville, VA 22903-0818}

\begin{abstract}

Isolated low-mass stars are formed, in the standard picture, from the 
collapse 
of dense cores condensed out of strongly magnetized molecular clouds. 
The dynamically collapsing inflow traps nearly half of the critical 
magnetic flux needed for the core support and deposits it in a small 
region surrounding the protostar. It has been argued previously that 
the deposited flux can slow down the  
inflow, allowing matter to pile up and settle along field lines into a 
magnetically supported, circumstellar disk. Here we show that the disk 
typically contains $\sim 10\%$ of the stellar mass, and that it could 
become self-gravitating under plausible conditions during the rapidly 
accreting, ``Class 0'' phase of star formation. Subsequent fragmentation 
of the self-gravitating, magnetically subcritical disk, driven by magnetic 
diffusion, could produce fragments of substellar masses, which collapse 
to form brown dwarfs and possibly massive planets. This scenario 
predicts substellar object formation at distances of order 100~AU 
from the central star, although orbital evolution is possible after 
formation. It may provide an 
explanation for the small, but growing, number of brown dwarf companions 
found around nearby stars by direct imaging. The relatively large 
formation distances make the substellar companions vulnerable to dynamic 
ejection, particularly in binary (multiple) systems and dense clusters. 
Those ejected may account for, at least in part, the isolated 
brown dwarfs and perhaps free-floating planetary mass objects.   
\end{abstract} 


\section{Introduction}
\label{intro}

The past several years have witnessed remarkable progress in the study 
of brown dwarfs (Basri 2000). To date, more than 200 of such objects
have been discovered. Most are found in isolation; only a small 
fraction orbit around stars, typically at separations of tens of AUs 
and beyond. Despite the impressive pace of discovery, investigation into 
their origins has only just begun. The isolated brown dwarfs could be 
formed in small multiple systems through cloud fragmentation like stars, 
but ejected through interaction with other (heavier) members 
before gaining enough mass for hydrogen burning (Reipurth \& Clarke 
2001). The brown dwarf companions could in principle be produced in 
(rotationally supported) circumstellar disks through gravitational 
fragmentation, although it is unclear whether 
the conditions for fragmentation can arise naturally during star
formation (Durisen 2001). In this Letter, we advance a new scenario 
for forming 
substellar companions, through the gravitational fragmentation of 
magnetically, rather than rotationally, supported disks. Such disks 
are expected to develop around protostars formed out of strongly 
magnetized clouds (Li \& McKee 1996; LM96 hereafter), 
as envisioned in the standard picture of isolated low-mass star 
formation (Shu, Adams \& Lizano 1987). 

Conceptually, the formation of magnetically supported disks parallels that
of the more familiar rotationally supported disks. Once gravity has
initiated the dynamic collapse of a rotating magnetized dense core of
molecular cloud, both the magnetic flux and the angular momentum
associated with rotation are trapped by the collapsing flow and carried to
the vicinity of the protostar. However both must eventually be stripped
almost completely from the matter before it enters the star; otherwise,
the stellar magnetic field strength and rotation rate would be many orders
of magnitude higher than actually observed. These are, respectively, the 
magnetic flux and angular momentum problem of star formation. The 
stripping of 
angular momentum is thought to take place primarily in a rotationally 
supported disk by friction between adjacent annuli of 
matter moving at different speeds. The mechanism for 
magnetic flux stripping is less clear. LM96 proposed that it 
occurs as the field 
lines decoupled from the stellar matter expand against the collapsing 
inflow, reversing the inward motion of charged particles and the 
magnetic field tied to them. The bulk neutral
material can still slip through, but only at a much reduced speed 
because of frequent collisions with the already stopped charged 
particles. The slowdown causes the infalling matter to pile up and 
settle along field lines into a flattened structure---a magnetically 
supported circumstellar disk (see also Ciolek \& K\"onigl 1998 and 
Contopoulos, Ciolek \& K\"onigl 1998). 

We show in this Letter that the magnetically supported disk could become 
self-gravitating under plausible conditions during the earliest, 
``Class 0'' phase of star formation, when the accretion rate is 
highest (\S~\ref{selfgrav}). The self-gravitating disk is magnetically 
subcritical, and fragmentation driven by magnetic diffusion could 
lead to substellar object formation (\S~\ref{diskfrag}). We discuss 
the expected characteristics of the substellar companions and their 
observational implications in \S~\ref{disscon}.

\section{Self-Gravitating Magnetically Supported Disks}
\label{selfgrav}

We consider the simplest case of axisymmetric non-self-gravitating disk 
and derive a condition for the disk to become self-gravitating around 
a single, isolated star $M_*$; possible nonaxisymmetric effects 
will be commented upon toward the end of the section.  We estimate 
the disk properties in a way similar to LM96, but include the magnetic 
compression, which controls the disk density, and the magnetic tension 
force, which dominates the magnetic pressure gradient in the radial
force balance: 
\begin{equation}
{GM_* / r^2}\approx {B_z B_r / 2\pi\Sigma}.
\label{e1}
\end{equation}
We show below that the disk column density $\Sigma$ is roughly independent 
of the cylindrical radius $r$. To counter the larger stellar gravity at 
a smaller radius, the field strength must increase inward, roughly as 
$r^{-1}$. For such a distribution, the radial and vertical field components 
are comparable on the disk (assuming a potential field outside as usual). 
They are   
\begin{equation}
B_z\approx B_r\approx {(2\pi G \Sigma M_*)^{1/2}/r}.
\label{e2}
\end{equation}
The magnetic flux enclosed within a disk of radius $R$ is 
\begin{equation}
\Phi_{\rm d}=\int_0^R2\pi B_z r dr\approx 2\pi R (2\pi G\Sigma M_*)^{1/2}.
\label{e3}
\end{equation}
Since this flux is mostly that stripped from the stellar mass $M_*$ (LM96), 
we have $\Phi_{\rm d}\approx \Phi_*= \epsilon (2\pi G^{1/2}) M_*$,
where $\epsilon$ ($< 1$) is the flux-to-mass ratio of the dynamically 
collapsing envelope that feeds the star-disk system in units of 
the critical value $2\pi G^{1/2}$. Eliminating $\Phi_{\rm d}$ from equation 
({\ref{e3}), we obtain the following estimate of the disk mass
\begin{equation}
M_{\rm d}\approx \pi R^2 \Sigma \approx 0.5 \epsilon^2 M_*,
\label{e5}
\end{equation}
which is simply the mass needed to ``weigh down'' the 
magnetic flux released from the star and prevent it from 
escaping. Even though the star-disk system remains magnetically 
supercritical, the disk (which contains a small fraction of the mass 
of the system but most of its flux) is subcritical, by a factor 
of $\sim 2/\epsilon$.   
Calculations of the evolution of strongly magnetized clouds 
driven by ambipolar diffusion suggest that 
$\epsilon$ is close to half (e.g., Nakamura \& Li 
2002). This value would yield a disk mass of $\sim 10\%$ of the stellar 
mass, or some $10^2$ Jupiter masses for solar-mass stars. Therefore, 
there appears to be enough matter in the disk to form one brown dwarf 
near the hydrogen burning limit, a few less massive brown dwarfs, or 
several massive planets. 

The mass density $\rho$ of the disk is determined by balancing the outward 
thermal pressure gradient in the vertical direction against the squeezing 
of the stellar tidal force, disk self-gravity, and the radial component of 
the magnetic field. It is easy to show that the magnetic contribution 
dominates and 
\begin{equation}
\rho\approx {B_r^2 / 8\pi a^2}\approx {G M_* \Sigma / 4 a^2 r^2},
\label{e9}
\end{equation}
where $a=0.188\; T_{10}^{1/2}$~km~s$^{-1}$ is the isothermal sound speed 
and $T_{10}\equiv T/10$~K. Equation ({\ref{e2}) was used to derive the 
second relation. 
The column density $\Sigma$ at any radius $r$ is related to the local 
mass accretion rate ${\dot M}$ and infall speed $V$ through 
$
\Sigma={{\dot M}/2\pi r V}.
$
Approximately, $V$ is the speed with which the bulk neutral matter 
slips across the field lines, which remain more or less fixed in 
space (LM96). The slippage speed can be written as a product of 
the total magnetic force per unit mass ($\approx G M_*/r^2$) and 
the magnetic coupling time $t_c$. The latter is related to the local
free-fall time $t_{\rm ff}=(3\pi/32G\rho)^{1/2}$ through 
$
t_c=t_{\rm ff}^2/t_B,
$
where the characteristic flux loss time $t_B$ from a magnetically 
supported cloud has been discussed extensively by Nakano and 
collaborators. According to Nakano, Nishi \& Umebayashi (2002), 
$t_B$ is a factor of $\sim 
10-10^2$ larger than $t_{\rm ff}$ up to a density of $\sim 
10^{10}-10^{11}$~cm$^{-3}$ for a cosmic ray ionization rate of $\sim 
10^{-17}-10^{-16}$~s$^{-1}$ and an MRN grain size distribution (see
their Fig.~3). It drops below $t_{\rm ff}$ beyond $\sim 10^{12}
$~cm$^{-3}$, as both ions and charged grains become decoupled 
from the field lines. The decoupling density depends somewhat
on the grain size distribution. It could be increased significantly 
due to rapid grain growth and/or sedimentation toward the disk 
midplane (Sano et al. 2000). We will concentrate on the part of 
the disk that is well coupled to the magnetic field (which makes 
mass pileup possible), and adopt $t_B=20\ t_{\rm ff}$ for a rough 
estimate. In this case, $t_c=t_{\rm ff}/20$ and the column 
density is given by 
\begin{equation}
\Sigma\approx 9\; {{\dot M}^2/ a^2 M_*}
=5\; {\dot m}^2 m_*^{-1} T_{10}^{-1}\ {\rm g~cm}^{-2}, 
\label{e10}
\end{equation}
where ${\dot m}$ is the accretion rate normalized by $10^{-5}\; {\rm M}
_\odot {\rm yr}^{-1}$ and $m_*$ the stellar mass in $M_\odot$. As
advertised, the column density is roughly independent of radius $r$ 
as long as the accretion rate does not vary much over the disk and
the disk is more or less isothermal. For a canonical dust opacity 
of 0.01~cm$^2$~g$^{-1}$, the disk is typically optically thin, making 
the isothermal approximation justifiable. 

In the absence of a strong magnetic field, a circumstellar disk becomes 
gravitationally unstable when the radius of a Jeans-mass clump is 
smaller than the Roche value. This condition roughly corresponds
to the Toomre's criterion (e.g., Shlosman \& Begelman 1989)
\begin{equation}
Q={a \Omega / \pi G\Sigma} \lesssim 1,
\label{e6}
\end{equation}
where $\Omega=(GM_*/r^3)^{1/2}$ is the Keplerian frequency, even though 
the disk is not rotationally supported. The presence of a strong 
magnetic field does not suppress the Jeans instability in the lightly 
ionized disk; it merely lengthens the growth timescale (Langer 
1978; see \S~\ref{diskfrag} for further discussion). Near the outer
edge of the disk 
\begin{equation}
R\approx \left({M_{\rm d}/ \pi \Sigma}\right)^{1/2}\approx 2.6\times
10^2 \epsilon_{_{0.5}} {\dot m}^{-1}  m_* T_{10}^{1/2}\ {\rm AU}, 
\label{e11}
\end{equation}
where most of the disk mass is located, the Toomre $Q$ parameter has 
the approximate value
\begin{equation}
Q_R\approx 0.8\; \epsilon_{_{0.5}}^{-3/2} {\dot m}^{-1/2} 
T_{10}^{3/4}, 
\label{e12}
\end{equation}
where $\epsilon_{_{0.5}}$ is the flux-to-mass ratio $\epsilon$ normalized 
by the typical value $1/2$. The most uncertain quantity in the above 
expression is the accretion rate 
${\dot m}$. If it is large enough (${\dot m}\gtrsim 1$), the disk could 
become self-gravitating (i.e., $Q_R < 1$). 

Low-mass protostars are thought to grow most rapidly during the so-called 
``Class 0'' phase (Andr{\'e}, Ward-Thompson \& Barsony 2000), which 
lasts for only a few times $10^4$ years. To accrete the bulk of the 
mass of Sun-like stars over such a short period of time, the accretion 
rate must be of order $10^{-5}$~M$_\odot$~yr$^{-1}$ or higher. Such high 
accretion rates are inferred from the observations of molecular outflows 
from Class 0 sources (Bontemps et al. 1996), and found in core collapse 
calculations (e.g., Tomisaka 1996; Li 1998). We conclude that during the
Class 0 phase of star formation the magnetically supported disk could 
become self-gravitating and thus prone to fragmentation into small pieces. 
Before discussing fragmentation, we comment on a possible complication.

The disk has a flux-to-mass ratio $B_z/\Sigma$ increasing toward the 
center, and should be unstable to (nonaxisymmetric) interchange 
instability (LM96; Ciolek et al. 1998) according to the criterion of 
Spruit \& Taam (1990), 
derived in the limit of frozen-in magnetic field and no self-gravity. 
Nonlinear developments of the instability could potentially drain the 
disk material faster than estimated above, and thus prevent the disk 
from becoming self-gravitating. However, in our case the magnetic 
field is not frozen in the matter and the disks relevant to companion 
formation are self-gravitating. Moreover, a realistic disk would have 
a significant amount of differential rotation and be magnetically 
linked to the surrounding medium, both of which tend to stabilize the 
instability (Spruit, Stehle \& Papaloizou 1995), as does the (likely 
differential) twisting of the disk field lines associated with magnetic 
braking (Sakurai 1989). Nonaxisymmetric calculations are needed to 
ascertain whether the combination of magnetic diffusion, self-gravity,
rotation, and field line linkage and twisting can suppress the 
instability altogether and, if not, to what extent nonlinear developments 
of the instability modify the disk structure. 

\section{Magnetic Diffusion-Driven Disk Fragmentation} 
\label{diskfrag}

The disk is magnetically subcritical. If frozen in matter, the magnetic 
field would prevent fragmentation completely (Nakano 1988). 
However, in a lightly ionized medium, the presence of a 
strong magnetic field does not change the criterion for fragmentation; 
rather, the fragmentation occurs on the (longer) magnetic diffusion, 
instead of dynamic, timescale (Langer 1978). The minimum mass of 
fragments will still be given by the (thermal) Jeans mass, which in a 
flattened geometry has the following approximate expression (Larson 1985)
\begin{equation}
M_{\rm J}={1.17 a^4 / G^2\Sigma} \approx 3\times 10^{-3} 
{\dot m}^{-2}m_* T_{10}^3\ {\rm M}_\odot.
\label{e13}
\end{equation}
It corresponds to $\sim 3$ times the Jupiter mass ($M_{\rm Jup}$) for typical 
parameters, which is in the planetary mass range. This typical Jeans 
mass is smaller than the oft-quoted opacity-limited minimum value 
($\sim 7\ M_{\rm Jup}$; Low \& Lynden-Bell 1976; Silk 1977), although 
in a disk geometry where radiation can escape more easily this minimum 
could be lower.  

Without magnetic field, a self-gravitating fragment of one Jeans mass 
will contract in a shorter time onto itself than toward the central 
star. This remains true even in the presence of a strong magnetic field, 
since both timescales are lengthened by the same factor, set by magnetic 
diffusion. The timescale for the magnetic field to diffuse out of a 
fragment of Jeans radius $R_{\rm J}=(M_{\rm J}/\pi\Sigma)^{1/2}$ is set 
roughly by $\tau\approx R_{\rm J}/ V$, where $V\approx t_c (G M_{\rm J}
/R_{\rm J}^2)$ is the slippage speed of neutral matter
across field lines driven by self-gravity. Adopting the same approximation 
for the magnetic coupling time $t_c=t_{\rm ff}/20$ as before, and 
computing the freefall time $t_{\rm ff}$ near the disk outer edge where
\begin{equation}
\rho_{_R}\approx 3\times 10^{-14} \epsilon_{_{0.5}}^{-2}{\dot m}^4
m_*^{-2} T_{10}^{-3}\ {\rm g~cm}^{-3}
\label{e150}
\end{equation}
[obtained from equations (\ref{e9}), (\ref{e10}) and (\ref{e11})], we 
find 
\begin{equation}
\tau \approx 3\times 10^4 \epsilon_{_{0.5}}^{-1} {\dot m}^{-2} 
m_* T_{10}^{3/2}\ {\rm yrs},
\label{e17}
\end{equation} 
which for typical parameters is comparable to the lifetime of Class 
0 phase. Once the fragment has become magnetically supercritical, it 
collapses dynamically from inside out to form one or more substellar 
objects, as in the standard scenario for forming isolated low-mass 
stars through ambipolar diffusion (Shu et al. 1987). The main 
difference lies in the density of the background, 
magnetically subcritical material, which is of order $\sim 10^{-20}
$~g~cm$^{-3}$ in the stellar case but $\sim 10^{-14}$~g~cm$^{-3}$ in the 
substellar case. The density, of course, sets the mass scale (for a $\sim 
10$~K gas). In the substellar case, the density is high enough that 
magnetic decoupling could be a potential problem; its complete 
treatment is hindered by uncertainties in the grain properties of 
the disk.  

We therefore arrive at the following plausible scenario for the production 
of substellar companions, based on the standard picture of isolated 
low-mass star formation. To avoid the famous ``magnetic flux problem'', 
most of the magnetic flux originally trapped by the collapsing core
material that 
enters the star must be stripped and deposited in a magnetically supported 
disk containing $\sim 10\%$ of the stellar mass. If the disk
material is not drained at too large a speed by interchange instability 
and/or magnetic decoupling, it could achieve a high enough surface density
during the Class 0 phase to become self-gravitating. 
Subsequent contraction of self-gravitating pockets 
of substellar masses across field lines, over a magnetic diffusion 
timescale comparable to the Class 0 lifetime, could produce 
magnetically supercritical fragments, which collapse to form substellar 
companions. This scenario does not involve rotation directly, and should 
not be affected by density waves, which tend to suppress gravitational
fragmentation of rotationally-supported disks (Durisen 2001). 

\section{Implications for Brown Dwarf Companions} 
\label{disscon}

If the substellar companions are formed from the fragmentation of 
magnetically supported disks, they should have initial masses between 
the opacity-limited Jeans mass, which is in the (upper) planetary mass 
range, and the disk mass, typically comparable to that of a massive brown 
dwarf. Further mass increase by accretion from the residual envelope 
is possible, but probably not by much because of rapid decline of the 
overall accretion rate after the Class 0 phase, as a result of, e.g., 
the powerful protostellar wind unbinding much of the envelope. If 
substantial mass accretion does occur, the substellar objects may be 
turned into (very) low mass stellar companions. 

A prediction of our scenario is that the substellar companions
should form at distances of a few hundred AUs from the central star, 
set by the typical size of the self-gravitating disk (cf. equation 
\ref{e11}). Interestingly, more than half of the brown dwarf companions 
detected so far have separations in the range of $\sim 50 - 300$~AU 
(see Reid et al. 2001 for a compilation). The orbits 
could shrink after formation for several reasons: first, the 
disk material out of which the objects form was mainly magnetically 
supported, and thus sub-Keplerian. 
%% Indeed, the slowdown of radial 
%% infall in the disk, coupled with the presence of a strong magnetic 
%% field, allows magnetic braking to operate more efficiently, which
%% could remove a large fraction of the disk angular momentum, similar 
%% to the case of magnetically subcritical clouds (Basu \& Mouschovias 
%% 1994). 
Second, the substellar objects could 
transport some of their orbital angular momentum to the residual 
envelope by raising tides in it. In addition, since the disk contains 
more than one Jeans mass, more than one object could form from its 
fragmentation, with the ejection of all but one as a likely 
outcome of gravitational interactions. The final orbit of the 
remaining object could tighten by a large factor and tends to be highly 
eccentric (Papaloizou \& Terquem 2001). Under extreme conditions,
these interactions could in principle explain the 
smaller separations of Gliese 86B ($\sim 19$~AU; Els et al. 2001) 
and HR 7672B ($\sim 14$~AU; Liu et al. 2002). To send a brown dwarf 
even closer in, say to the $\sim 3$~AU distance of HD 168443c (Marcy et al. 
2001), would be even more difficult, consistent with the well known 
dearth of brown dwarf companions within $\sim 4$~AU of solar-type 
FGK stars---the so-called ``brown dwarf desert''. For the same reason, 
our scenario may produce (massive) extrasolar giant planets on 
relatively wide 
orbits of tens to hundreds of AUs, but unlikely those within a few 
AUs of their host stars detected through radial-velocity surveys. 
These close-in planets are presumably 
formed in rotationally supported disks, possibly through the 
conventional core-nucleation mechanism. Their orbital evolution
may be affected, however, by the presence of substellar companions 
at larger distances. 

The substellar companions can also increase their orbital separations
after formation, through gravitational interaction with the Keplerian 
disks they enclose. If formed in binary (or multiple) systems or dense 
clusters, they would likely be ejected through dynamic interactions, 
given their relatively large formation distances from the central star 
(Reipurth \& Clarke 2001), although some may remain bound on wider 
orbits. Such interactions could in principle explain the 
the few thousand AU separations of the brown dwarf companions found in 
the 2MASS survey and follow-up observations, including Gl 570D, Gl 417B, 
Gl 584C, Gl 337C, and possibly HD 89744B and Gl 618.1B (Kirkpatrick et al. 
2001; Wilson et al. 2001). Those ejected 
could account for, at least in part, the isolated brown dwarfs and possibly 
the free-floating planetary mass objects that may have been uncovered in
$\sigma$ Orionis (Zapatero Osorio et al. 2000) and elsewhere.

An esthetically appealing feature of our scenario is that the brown dwarf 
companions are produced in essentially the same manner as the isolated 
low-mass stars envisioned in the standard picture, through the magnetic 
diffusion-driven fragmentation of 
a strongly magnetized, self-gravitating medium. Two common products of 
the fragmentation in the stellar case are circumstellar disks and 
binaries. By analogy, we expect disks and binaries to result from the 
fragmentation in the substellar case as well. A surrounding disk could 
explain the infrared excess observed in the young brown dwarf in the GG 
Tau quadruple system (White et al. 1999), and brown dwarf pairs have been 
detected (Basri 2000). 
Dynamical ejection from the $10^2$~AU-scale distance from a 
Sun-like low-mass star should leave intact the inner part of the disk 
(within a few AUs of the brown dwarf) or tightly bound pairs. The 
former is consistent with the inference of disks around young isolated 
brown dwarfs (Muench et al. 2001), and the latter with the dearth of 
isolated brown dwarf binaries with separations greater than $\sim 10$~AU 
(Reid et al. 2001).

I thank G. Laughlin, C. McKee, F. Nakamura, R. Nishi, C. Terquem and
the referee for helpful comments.


\begin{thebibliography}{}

%\bibitem[Basu(2000)]{SBasu00}
%      Basu, S. 2000, \apj, 540, 103

\bibitem[]{}
Andr\'e, P., Ward-Thompson, D. \& Barsony, M. 2000, in 
Planets and Protostars IV, eds. V. Mannings, A. Boss, 
\& S. Russell (Arizona: Univ of Arizona Press), p59

%\bibitem[]{}
%Basu, S. 1997, ApJ, 485, 240
\bibitem[]{}
Basri, G. 2000, ARAA, 38, 485

%\bibitem[]{}
%Basu, S. \& Mouschovias, T. Ch. 1994, ApJ, 432, 720 
%
\bibitem[]{}
Bontemps, S., Andr\'e, P., Tereby, S. \& Cabrit, S. 1996, AA, 311, 858

%\bibitem[]{}
%Burgasser, A. J., Kirkpatrick, J. D., et al. 2000, ApJ, 531, L57
%
\bibitem[]{}
Ciolek, G. E. \& K\"onigl, A. 1998, ApJ, 504, 257

\bibitem[]{}
Contopoulos, I., Ciolek, G. E. \& K\"onigl, A. 1998, ApJ, 504, 247

\bibitem[]{}
Durisen, R. H. 2001, in The Formation of Binary Stars, eds. H. Zinnicker
\& R. Mathieu, IAU200, p381

\bibitem[]{}
Els, S. G., Sterzik, M. F., Marchis, F. et al. 2001, AA, L1

%\bibitem[]{}
%Foster, P. N. \& Chevalier, R. A. 1993, ApJ, 416, 303
%
%\bibitem[Galli et al.(2001)]{DGalli01}
%Galli, D., Shu, F. H., Laughlin, G., \& Lizano, S. 2001, \apj,
%	551, 367

%\bibitem[]{}
%Henriksen, R. N., Andr\'e, P. \& Bontemps, S. 1997, AA, 323, 549

\bibitem[]{}
Kirkpatrick, J. D., Dahn, C. C., et al. 2001, AJ, 121, 3235

%\bibitem[]{}
%Lane, B. F., Zapatero Osorio, M., Britton, M., Martin, E. L. \& Kulkarni,
%S. R. 2001, ApJ, 560, 390
%

\bibitem[Langer(1978)]{WLanger78}
      Langer, W.D. 1978, \apj, 225, 95

\bibitem[]{}
Larson, R. B., 1985, MNRAS, 214, 379
%
%\bibitem[]{}
%------, 2002, MNRAS, in press

\bibitem[]{}
Li, Z.-Y. 1998, ApJ, 493, 850

\bibitem[]{}
Li, Z.-Y. \& McKee, C. F. 1996, ApJ, 464, 373 (LM96)

\bibitem[]{}
Liu, M. C., Fisher, D., Graham, J. R., Lloyd, J. P., Marcy, 
G. W. \& Butler, R. P. 2002, ApJ, in press

\bibitem[]{}
Low, C. \& Lynden-Bell, D. 1976, MNRAS, 176, 367

\bibitem[]{}
Marcy, G. W., Bulter, R. P., Vogt, S. S. et al. 2001, ApJ, 554, 418

%\bibitem[]{}
%Marcy, G. W., Cochran, W. D. \& Mayor, M. 2000, in 
%Planets and Protostars IV, eds. V. Mannings, A. Boss, 
%\& S. Russell (Arizona: Univ of Arizona Press), p1285
%
%\bibitem[]{}
%Mouschovias, T. Ch. \& Ciolek, G. E. 1999, in The Origins of Stars and 
%Planetary Systems, eds C. Lada \& N. Kylafis (Dordrecht: Kluwer), p305

\bibitem[]{}
Muench, A. A., Alves, J., Lada, C. J. \& Lada, E. A. 2001, 558, L51

%\bibitem[Myers et al.(1991)]{PMyers91}
%Myers, P. C., Fuller, G. A., Goodman, A. A., \& Benson,
%	P. J. 1991, \apj, 376, 561
%
%\bibitem[]{}
%Nakajima, T., Oppenheimer, B. R., Kulkarni, S. R. et al. 1995, Nature, 378, 
%463
%
\bibitem[]{}
Nakamura, F. \& Li, Z.-Y. 2002, ApJ, 566, L101 

\bibitem[]{}
Nakano, T. 1988, PASJ, 40, 593

\bibitem[]{}
Nakano, T., Nishi, R. \& Umebayashi, T. 2002, ApJ, in press 
(astro-ph/0203223)

%\bibitem[]{}
%Nishi, R., Nakano, T. \& Umebayashi, T. 1991, ApJ, 368, 181
%

\bibitem[]{}
Papaloizou, J. C. B. \& Terquem, C. 2001, MNRAS, 325, 221

%\bibitem[]{}
%Potter, D., Martin, E. L., Cushing, M. C. et al. 2002, ApJL, in press
%
\bibitem[]{}
Reid, I. N., Gizis, J. E., Kirkpatrick, J. D. \& Koerner, D. W. 2001, AJ,
121, 489

\bibitem[]{}
Reipurth, B., \& Clarke, C. 2001, ApJ, 122, 432

\bibitem[]{}
Sakurai, T. 1989, Solar Physics, 121, 347

\bibitem[]{}
Sano, T., Miyama, S. M., Umebayashi, T., \& Nakano, T. 2000, ApJ, 543, 486

\bibitem[]{}
Shlosman, I. \& Begelman, M. C., 1989, ApJ, 341, 685

%\bibitem[]{}
%Shu, F. H. 1995, in Molecular Clouds and Star Formation, 
%eds C. Yuan \& J. You
%(Singapore: World Scientific), p97

\bibitem[]{}
Shu, F. H., Adams, F. C. \& Lizano, S. 1987, ARAA, 25, 23

\bibitem[]{}
Silk, J. 1977, ApJ, 214, 152

\bibitem[]{}
Spruit, H. C., Stehle, R. \& Papaloizou, J. C. B. 1995, MNRAS, 275, 1223

\bibitem[]{}
Spruit, H. C. \& Taam, R. E. 1990, AA, 229, 475

\bibitem[]{}
Tomisaka, K. 1996, PASJ, 48, L97

%\bibitem[]{}
%Umebayashi, T. \& Nakano, T. 1990, MNRAS, 243, 103
%
%\bibitem[]{}
%van der Tak, E. F. S. \& van Dishoeck, E. F. 2000, AA, 358, L79
%
\bibitem[]{}
White, R. J., Ghez, A. M., Reid, I. N. \& Schultz, G. 1999, ApJ, 520, 811

\bibitem[]{}
Wilson, J. C., Kirkpatrick, J. D., Gizis, J. E. et al. AJ, 122, 1989

\bibitem[]{}
Zapatero Osorio, M. R. et al. 2000, Science, 290, 103

\end{thebibliography}
\end{document}